\documentclass[aps,pre,twocolumn,showpacs,groupedaddress,amsmath,amssymb]{revtex4}

\usepackage{graphicx,psfrag}

\def\BibTeX{{\rm B\kern-.05em{\sc i\kern-.025em b}\kern-.08em
    T\kern-.1667em\lower.7ex\hbox{E}\kern-.125emX}}

\begin{document}

\title{Anticipatory synchronization with variable time delay and reset}

\author{G. Ambika}
\email{g.ambika@iiserpune.ac.in}
\affiliation{Indian Institute of Science Education and Research, Pune-411 021, India}
\author{R. E. Amritkar}
\email{amritkar@prl.res.in}
\affiliation{Physical Research Laboratory, Ahmedabad-380 009, India}

\begin{abstract}
A method to synchronize two chaotic systems with anticipation or lag, coupled in the drive response mode is proposed. The coupling involves variable delay with three time scales. The method has the advantage that synchronization is realized with intermittent information about the driving system at intervals fixed by a reset time. The stability of the synchronization manifold is analyzed with the resulting discrete error dynamics. The numerical calculations in standard systems like the R\"ossler and Lorenz systems are used to demonstrate the method and the results of the analysis.
  
\end{abstract}


\maketitle

\section{\label{sec1}INTRODUCTION}

The synchronization of uni-directionally coupled chaotic systems have been studied reasonably well in the past few years \cite{pec,tre,amr}. The synchronization state in such cases can be phase, lag, generalized or complete depending upon the strength of the coupling \cite{boc,cor}. Recently synchronization of systems via coupling with a time delay, which presumably takes care of the finite propagation times, switching speeds and memory effects have been reported \cite{pyr,zho,cho,bun}. Such studies relate to a variety of diverse phenomena like chirping of crickets, neural networks, automatic steering and control and coupled phase locked lasers \cite{ste,ern,sha}. When the coupling is not isochronous with the system dynamics, it is possible to realize retarded (delay), complete and anticipatory synchronization of chaotic systems. Moreover, synchronization in such cases reveal many novel phenomena like parametric resonance\cite {zho1}, multi-stable phase clustering \cite{par,wan1,seth,nak}, amplitude death etc \cite{kon,pra,ram}. An interesting aspect of such delay induced synchronization, that has attracted lot of attention, is that the driven system can anticipate the dynamics of the driver \cite{vos1,wan2}. The maximum possible anticipation time is reported to be enhanced considerably by using an array or ring of such systems \cite {vos2,wan3,cis}. Experimental verification of anticipatory and retarded synchronization is reported in electronic circuits as well as semiconductor lasers with delayed optoelectronic feed back \cite {hail,tan}. In all these studies,  the delay time in the coupling, once chosen, remains constant as the system evolves.  

The synchronization of chaotic systems, in general, has attracted great attention due its potential application in secure communication \cite {he,li,wan4}. However the use of low dimensional systems in this context is found to be insecure due to the ease of reconstruction from the transmitted signal \cite{vai,sho}. Therefore recently, chaos synchronization in high dimensional systems, especially systems with an inherent time delay, has been proposed as a better alternative \cite{pen,uda,goe,gho,yao}.

In this paper we propose a method of delay/anticipatory synchronization with coupling involving variable time delay.  Here, the synchronization can be realized with limited information about the driver via occasional contacts or feedbacks at specific intervals. This makes the method highly cost effective and can be applied to cases where the signal transmission from driver is slow or intermittent. This is achieved by using a variable delay in the coupling that is reset at definite intervals. The dynamics  then evolves under three additional time scales, the delay $\tau_1$, the anticipatory time $\tau_2$ and the reset time $\tau$. Unlike the case of fixed delay, the resetting mechanism makes the error dynamics discrete and it is
possible to carry out an approximate analytic analysis. The analysis gives the maximum $\tau_2$ for a given $\tau$. This also fixes  the regions of stability in the parameter plane of coupling and delay. The method is demonstrated for standard systems like R\"ossler and Lorenz.

\section{\label{sec:level1}SYNCHRONIZATION WITH VARYING DELAY AND RESET}
\subsection{Model system}
Consider a dynamical system $x$ of dimension $n$ that drives an identical system $y$. We choose a simple coupling term of the linear difference type but with the drive variable delayed by $\tau_1$ and the driven variable delayed by $\tau_2$. Thus, the dynamics is given by 
\begin{subequations}
\label{system_def}
\begin{eqnarray}
    \dot{x} & = & f(x) 
\label{system_def_x}\\ 
    \dot{y} & = & f(y) + \epsilon \sum_{m=0}^{\infty} \Gamma \left(x_{t_1}-y_{t_2}\right) \chi_{(m\tau,(m+1)\tau)}
 \label{system_def_y}
\end{eqnarray}
\end{subequations}
where $x_{t_1} = x(t-t_1)$, $y_{t_2} = y(t-t_2)$, $\tau$ is the resetting time and $\chi_{(t',t'')}$ is an 
indicator function such that
$\chi_{(t',t'')} = 1 \; \textrm{for} \; t' \leq t \leq t''$ and
zero otherwise. Here, $\Gamma=[\Gamma_1,\Gamma_2,\ldots,\Gamma_n]^T$ is a constant vector specifying the coupling between the components of $x$ and $y$. In numerical simuations we take only one coponent of $\Gamma$ to be nonzero. Both the delays $t_1$ and $t_2$ depend on time and 
we choose this dependence as
\[ t_i = \tau_i + t - m\tau, \; \; i=1,2. \] Thus, $t-t_i = m\tau - \tau_i$.
As the two systems evolve, $t_1$ and $t_2$ also evolve with the same time scale and the coupling term uses the same value of both variables 
$x_{t_1}$ and $y_{t_2}$ during each resetting time interval $\tau$, 
i.e. the coupling term is constant for the time interval $\tau$. In each time interval $\tau$, the initial values of the delays $t_1$ and $t_2$ are $\tau_1$ and $\tau_2$ respectively. The delays increase linearly with time upto values $\tau_1+\tau$ and $\tau_2+\tau$ and then they are reset for the next interval. As a 
consequence, the coupling requires the variable of the drive system only at 
discrete time intervals of $\tau$. We also note that $t_1-t_2 = \tau_1-\tau_2$ for all $t$.

\subsection{Synchronization Manifold}
Synchronization manifold for the coupled systems (\ref{system_def}) is
defined by
$y(t-\tau_2) = x(t-\tau_1) \; {\textrm or} \;
y(t) = x(t-\tau_1+\tau_2)$
Thus, we  can get all the following three possibilities \cite{senthil}. 
(1) If $\tau_1-\tau_2 > 0$, we can get delay or lag synchronization with 
$\tau_1-\tau_2$ as the lag time. (2) If $\tau_1-\tau_2 < 0$, we can get 
anticipatory synchronization with $\tau_2-\tau_1$ as the anticipation time. 
(3) If $\tau_1-\tau_2 = 0$, we can get equal time synchronization.

As an illustration of this, we take the standard R\"ossler oscillator in the chaotic state as the driver described by the equations
\begin{eqnarray}
	\dot{x_1} & = & - x_2 - x_3 \nonumber \\
	\dot{x_2} & = & x_1 + a x_2 
\label{rossler} \\
	\dot{x_3} & = & b + x_3(x_1 - c). \nonumber
\end{eqnarray}
This is coupled to an identical system through the coupling scheme given in Eq.~(\ref{system_def}). Only $x_1$ and $y_1$ are coupled, i.e. $\Gamma = [1,0,0]^T$. Taking the parameter values $a= 0.15, b=0.2$ and $c= 10.0$, both the systems are evolved from random initial conditions using Runge Kutta algorithm with a time step 0.01 for 2000 units of time. With $\tau =0.10$  and the coupling strength $\epsilon = 0.4$, the resulting  time series obtained for $\tau_1$=0.84 and $\tau_2$ = 0.02 is plotted in Fig.1.a. Here the response system y(t)(dashed line) lags behind the driver x(t)(solid line) by $\tau_2 - \tau_1$. Fig.1.b shows the same for $\tau_1$= 0.02 and $\tau_2$=0.84 where y(t) anticipates x(t) with the same time shift. The degree of synchronization with the corresponding time shift can be quantified using the similarity function defined as 
\begin{equation}
S^2(T) = \frac{<[y_1(t)-x_1(t+T)]^2>}{\sqrt{<x_1^2 (t)><y_1^2 (t)>}}
\end{equation}
Figs.1.c and 1.d show $S^2(T)$ computed for  different values of $T$. The minimum occurs at 0.82, i.e. $T=|\tau_1-\tau_2|$, indicating synchronization with delay or anticipation of 0.82 time units.

It should be noted that the delay time $\tau_1$ is not
of much significance in the error dynamics, since the time scale of the drive system can be
linearly shifted by $\tau_1$. This point will become clear when we do
the stability analysis in the next section.

\begin{figure}
\includegraphics[width=0.95\columnwidth]{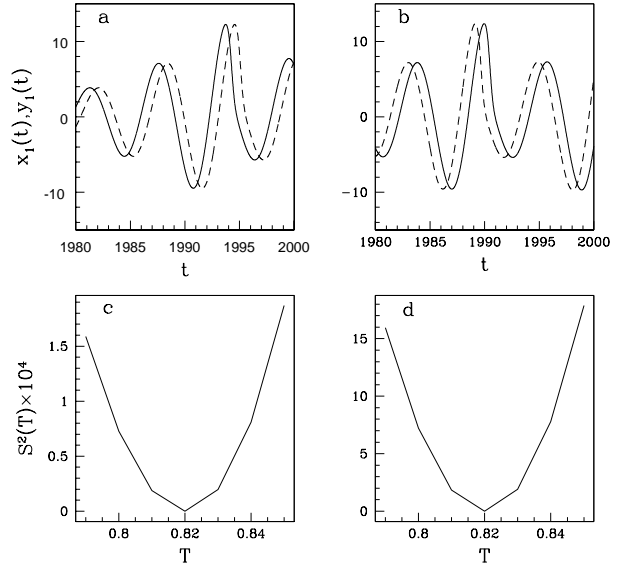}%
\caption{\label{del-ant-ros.eps}The simulated time series of two R\"ossler systems coupled through the scheme in Eq.~(\ref{system_def}). In (a) the case of delay synchronization is shown with a delay of 0.82 units between the $x_1(t)$(solid line) and $y_1(t)$(dashed line). (b) is a case of anticipatory synchronization when $y_1(t)$ anticipates $x_1(t)$ by the same units. The similarity function $S^2(T)$ corresponding to both these cases are shown in (c) and (d) respectively. 
}
\end{figure}

\section{\label{sec2}LINEAR STABILITY ANALYSIS}

The dynamics of the system in Eq.~(\ref{system_def}) involves three time scales in addition to its inherent scale. Define the transverse system by the variable $\Delta = y - x_{\tau_1-\tau_2}$. Its dynamics in 
linear approximation can be derived from Eq.~(\ref{system_def}) as 
\begin{eqnarray}
 \dot{\Delta} = f^{'}(x_{\tau_1-\tau_2}) \Delta - 
\epsilon \sum_{m=0}^{\infty} \chi_{(m\tau,(m+1)\tau)} \Delta_m 
\label{linear_stability}
\end{eqnarray}
where $\Delta_{m} = \Delta(t-t_2) = \Delta(m\tau-\tau_2)$ and we take coupling in all components of $x$ and $y$, i.e. $\Gamma=[1,1\ldots,1]^T$. Thus,
$\Delta_m$ is a constant in each time interval 
$m\tau \leq t < (m+1) \tau$.
We note that $\tau_1$ enters only through the Jacobian term $f^{'}$ and can be eliminated by shifting the time scale of the drive system linearly and redefining $\tau_2$ suitably. Hence, as noted in the previous section, $\tau_1$ is not very 
significant for the stability analysis. The fixed point $\Delta=0$ corresponds to the lag/anticipatory synchronized state.

In general, it is not possible to solve Eq.~(\ref{linear_stability}). However, we can
approximate the equation by replacing Jacobian $f^{'}$ by some effective time 
average Lyapunov exponent $\lambda$ (only the real part is required).
\begin{equation}
 \dot{\Delta} = \lambda \Delta - \epsilon \sum_{m=0}^{\infty}  \chi_{(m\tau,(m+1)\tau)} \Delta_m
\label{linear_problem}
\end{equation}
In the following analysis we assume $\lambda$ to be positive. The results
can be easily extended to $\lambda <0$ (see Appendix \ref{AppC}).

>From the numerical analysis presented in the next section, it appears that the approximation of replacing $f^{'}$ by an effective $\lambda$ is reasonable for small values of $\tau_2$. We need a larger value of $\lambda$ for large $\tau_2$.

In the interval $m\tau \leq t < (m+1) \tau$, the solution of Eq.~(\ref{linear_problem}) is
\begin{equation}
 \Delta = \alpha \Delta_m + C_m e^{\lambda t}
\label{sol-Cm}
\end{equation}
where $\alpha = \epsilon / \lambda$ is the normalized dimensionless coupling constant, and $C_m$ is an integration constant.

\subsection{$0 \leq \tau_2 \leq \tau$}
Let us first consider the case $0 \leq \tau_2 \leq \tau$. For $t = (m+1)\tau-\tau_2$, $\Delta = \Delta_{m+1}$. Thus, eliminating the integration constant, Eq.~(\ref{sol-Cm}) gives
\begin{equation}
 \Delta = \alpha \Delta_m + (\Delta_{m+1} - \alpha \Delta_m) e^{\lambda (t-(m+1)\tau + \tau_2)}
\label{sol-m}
\end{equation}
For $(m-1)\tau \leq t \leq m\tau$
we have
\begin{equation}
\Delta = \alpha \Delta_{m-1} + (\Delta_m - \alpha \Delta_{m-1}) e^{\lambda (t-m\tau + \tau_2)}
\label{sol-m-1}
\end{equation}
Matching the solutions (\ref{sol-m}) and~(\ref{sol-m-1}) at $t=m\tau$, and
simplifying, we get the following recursion relation
\begin{subequations}
\label{rec-tau2}
\begin{eqnarray}
 \Delta_{m+1} & = & \alpha (1-e^{\lambda(\tau-\tau_2)} + \frac{1}{\alpha}
 e^{\lambda \tau}) \Delta_m
\nonumber \\
& &    - \alpha e^{\lambda \tau}(1-e^{-\lambda \tau_2}) \Delta_{m-1} 
\label{rec-tau2-a} \\
 & = & a \Delta_m - b \Delta_{m-1} 
\label{rec-tau2-b}
\end{eqnarray}
\end{subequations}
where
\begin{subequations}
\label{ab}
\begin{eqnarray}
 a & = & \alpha (1-e^{\lambda(\tau-\tau_2)}) + e^{\lambda \tau},
\label{a} \\
b & = & \alpha e^{\lambda \tau}(1-e^{-\lambda \tau_2}).
\label{b}
\end{eqnarray}
\end{subequations}
We can write Eq ~(\ref{rec-tau2-b}) as a 2-d map in matrix form as,
\begin{equation}
 \left( \begin{array}{c}
 \Delta_{m+1} \\ \Delta_m
\end{array} \right) = 
\left( \begin{array}{cc}
 a & -b \\ 1 & 0
\end{array} \right)
\left( \begin{array}{c}
 \Delta_m \\ \Delta_{m-1}
\end{array} \right)
\label{2dmap}
\end{equation}
The eigenvalue equation for the Jacobian matrix is 
\begin{equation}
 \mu^2 - a \mu  + b =0
\label{eq-ev-2d}
\end{equation}
with the  solutions 
\begin{eqnarray}
 \mu_{\pm} & = & \frac{1}{2}(a \pm \sqrt{a^2-4b})
 \label{ev-a}
\end{eqnarray}
The synchronized state, $\Delta=0$, is stable if both the solutions
satisfy $|\mu_{\pm}| < 1$. The detailed analysis of the stability conditions
is given in Appendix \ref{AppA}. 

Fig.~\ref{tau2-ep} shows the stability region in
$\tau_2/\tau-\alpha$ plane. The lower limit of stability is always
$\alpha_l=1$. For smaller values of $\tau_2$ ($\tau_2 \leq \tau_{2p}$), the upper limit
of stability is given by (Eq.~(\ref{alpha-case2}))
\begin{equation}
 \alpha_u = \frac{e^{\lambda \tau}+1}{2e^{\lambda (\tau- \tau_2)} - e^{\lambda \tau} - 1}
\label{alphau1}
\end{equation}
while for larger values of $\tau_2$ ($\tau_{2p} \leq \tau_2 \leq \tau$) it is given by (Eq.~(\ref{alpha-case3}))
\begin{equation}
 \alpha_u = \frac{e^{-\lambda \tau}}{1-e^{-\lambda \tau_2}}
\label{alphau2}
\end{equation}
The maximum value of $\alpha_p$ is given by the intersection of the two curves~(\ref{alphau1}) and~(\ref{alphau2}).
\begin{equation}
 \alpha_p = \frac{3+e^{\lambda \tau}}{e^{\lambda \tau}-1}
\label{alphap}
\end{equation}
The corresponding $\tau_{2p}$ value is given by
\begin{equation}
 \tau_{2p} = \frac{\alpha_p (\alpha_p+3)}{(\alpha_p+1)^2}
\label{tau2p-alphap}
\end{equation}

\begin{figure}
\includegraphics[width=0.9\columnwidth]{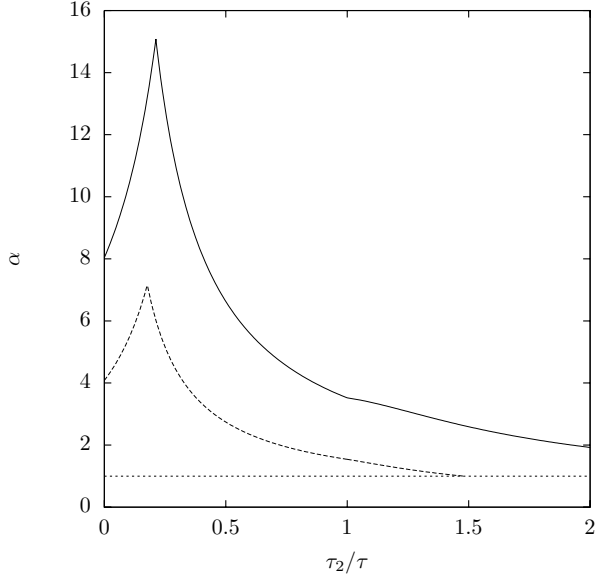}%
\caption{\label{tau2-ep}This figure shows the stability region
of the synchronized state ($\Delta=0$ solution of Eq.~(\ref{rec-tau2}))
in the $\tau_2/\tau - \alpha$ plane. The solid line is for  $\lambda \tau = 0.25$ and the dashed line is for $\lambda \tau = 0.5$. The
lower limit of stability is 
$\alpha_l=1$ (dotted line). For smaller values of $\tau_2 \leq \tau_{2p}$, the upper limit
of stability is given by (Eq.~(\ref{alphau1}))
while for larger values of $\tau_2$ ($\tau_{2p} \leq \tau_2 \leq \tau$) it is given by (Eq.~(\ref{alphau2})). The peak values are ($0.0530\ldots/0.25\ldots=0.212 ,15.083\ldots$) for $\lambda\tau=0.25$ and ($0.088\ldots/0.5=0.176\ldots,7.169\ldots$)for $\lambda\tau=0.5$ (see Eqs.~(\ref{alphap}) and~(\ref{tau2p-alphap})). For $\tau \leq \tau_2 \leq 2 \tau$ the upper limit of stability is given by Eq.~(\ref{alphau-k1}).The maximum value of $\lambda\tau_2$ is $0.74$ for $\lambda\tau = 0.5$.
}
\end{figure}

We also obtain $\tau_{2max}$, the maximum allowed value of $\tau_2$ for the stability of the synchronized state (Eq.~(\ref{tau2max-0})). A general expression for $\tau_{2max}$ is obtained in the next subsection (Eq.~(\ref{tau2max-k})).

\subsection{$\tau_2 > \tau$}
Let $\tau_2 = k\tau + \tau_2^{'}, \; k = 0,1,\ldots$ where $\tau_2^{'} < \tau$. 
Consider the solution~(\ref{sol-Cm}) in the interval $m\tau \leq t \leq (m+1)\tau$. Then for $t= (m+1)\tau - \tau_2^{'} = (m+k+1)\tau - \tau_2$, we get 
\begin{eqnarray} 
 \Delta_{m+k+1} & = & \alpha \Delta_m + C_m e^{\lambda (m+1) \tau - \lambda \tau_2^{'}}
\end{eqnarray}
Hence Eq.~(\ref{sol-Cm}) becomes
\begin{equation}
 \Delta = \alpha \Delta_m + (\Delta_{m+k+1} - \alpha \Delta_m) e^{\lambda (t-(m+1)\tau + \tau_2^{'})}
\label{sol-mk}
\end{equation}
For $(m-1)\tau \leq t \leq m\tau$
we have
\begin{equation}
\Delta = \alpha \Delta_{m-1} + (\Delta_{m+k} - \alpha \Delta_{m-1}) e^{\lambda (t-m\tau + \tau_2^{'})}
\label{sol-mk-1}
\end{equation}
Equating the solutions (\ref{sol-mk-1}) and~(\ref{sol-mk}) for $t=m\tau$,
and simplifying we get the following recursion relation
\begin{eqnarray}
 \Delta_{m+k+1} = && e^{\lambda \tau} \Delta_{m+k} - \alpha (e^{\lambda(\tau-\tau_2^{'})} -1) \Delta_m\nonumber\\
 &&- \alpha e^{\lambda \tau}(1-e^{-\lambda \tau_2^{'}}) \Delta_{m-1} 
\label{rec-tau2'} 
\end{eqnarray}
This gives a map of dimension $k+2$. In matrix form, the map can be expressed as
\begin{eqnarray}
 \left( \begin{array}{c}
 \Delta_{m+k+2} \\ \Delta_{m+k+1} \\ \Delta_{m+k} \\ \vdots \\ \Delta_m 
\end{array} \right) & = &
\left( \begin{array}{ccccc}
c & 0 & \ldots & b_1 & b_0 \\ 
1 & 0 & \ldots & 0 & 0 \\ 0 & 1 & \ldots & 0 & 0 \\
\vdots & & & & \vdots \\ 0 & 0 & \ldots & 1 & 0
\end{array} \right)
\left( \begin{array}{c}
 \Delta_{m+k+1} \\ \Delta_{m+k} \\ \Delta_{m+k-1} \\ \vdots \\ \Delta_{m-1}
\end{array} \right) \nonumber \\
& & \label{kdmap}
\end{eqnarray}
where $c=e^{\lambda \tau}$, $b_1=\alpha (e^{\lambda(\tau-\tau_2^{'})} -1)$
and $b_0=\alpha e^{\lambda \tau}(1-e^{-\lambda \tau_2^{'}})$. The eigenvalue 
equation is 
\begin{equation}
\mu^{k+2} - c \mu^{k+1} + b_1 \mu + b_0 = 0
\label{kth}
\end{equation}
For $k=0$, the map of Eq.~(\ref{kdmap}) reduces to the 2d-map of Eq.~(\ref{2dmap}).
In general the behavior of the largest magnitude $\mu$ is as shown in
Fig.~\ref{mu-alpha}b, i.e. the stability range is from $\alpha_l=1$ till
the complex $\mu$ has magnitude one.

\subsubsection{$k=1, \; {\textrm i.e.} \; \tau \leq \tau_2 \leq 2\tau$}
For $k=1$, we have a 3d-map. The eigenvalue equation (\ref{kth}) becomes
\begin{equation}
\mu^3 - c \mu^2 + b_1 \mu + b_0 = 0
\label{cubic-1}
\end{equation}
The lower stability limit is $\alpha_l=1$.
The upper stability limit can be obtained by noticing that when the magnitude of
the imaginary $\mu$ becomes one, the two imaginary eigenvalues can be written as $\mu = e^{\pm i \theta}$ and the above equation has a factor $\mu^2 - 2 \cos(\theta) \mu + 1$. This gives the condition
\begin{equation}
b_0^2 + c b_0 +b_1 -1 = 0 
\end{equation}
Using this we get
a quadratic equation for $\alpha$. 
\begin{equation}
 a_2 \alpha^2 + a_1 \alpha  - 1 =0
\label{eq-alpha-1}
\end{equation}
where $a_1 = e^{2 \lambda \tau} ( 1 - e^{-\lambda \tau_2^{'}}) + e^{\lambda(\tau-\tau_2^{'})} -1$ and $a_2 = e^{2 \lambda \tau}(1-e^{-\lambda \tau_2^{'}})^2$. One solution of this equation gives 
the upper stability limit for $\alpha$.
\begin{equation}
 \alpha_u = \frac{1}{2 a_2}(-a_1 + \sqrt{a_1^2 +4 a_2})
\label{alphau-k1}
\end{equation}
This upper stability limit is shown in Fig.~\ref{tau2-ep} for $\tau \leq \tau_2 \leq 2\tau$. 

We can also obtain $\tau_{2max}^{'}$, the maximum value of $\tau_{2}^{'}$
for which synchronization is possible. This happens when there is always an eigenvalue with magnitude greater than one, i.e. when $\alpha_l = \alpha_u =1$. By putting $\alpha_u=\alpha_l=1$ in Eq.~(\ref{eq-alpha-1}), we get $a_1 + a_2 = 1$. However a better condition is obtained if we
note that for $\alpha_u=\alpha_l=1$, Eq.~(\ref{cubic-1}) has two 
degenerate solutions $\mu=1$, i.e. Eq.~(\ref{cubic-1}) has a factor 
$\mu^2 - 2 \mu +1$. This gives the conditions $b_0 = 2- c$ and 
$1-b_1 = 2(2-c)$. First condition gives
\begin{eqnarray}
 \lambda \tau_{2max} & = & \lambda \tau + \lambda \tau_{2max}^{'}
\nonumber \\
 & = & \lambda \tau - \ln 2 - \ln(1-e^{-\lambda \tau})
\label{tau2max-1}
\end{eqnarray}

\subsubsection{General $k$}
\begin{figure}
\includegraphics[width=0.9\columnwidth]{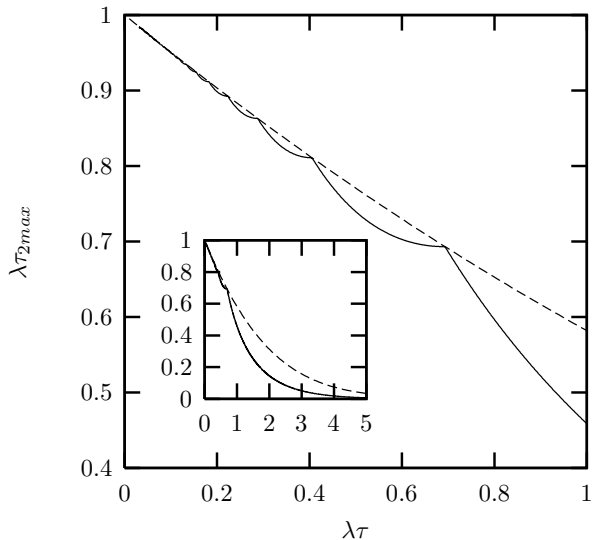}%
\caption{\label{tau-tau2max}The figure plots the maximum $\tau_{2max}$ as
a function of $\lambda \tau$ (solid line). Here, $\tau_{2max}, \; k=0,1,2,\ldots$ is given by Eq.~(\ref{tau2max-k})) and for each $k$ the range of $\tau$ is
$(\tau_k,\tau_{k+1})$. The dashed
curve passes through the values $\tau_{2max} = k\tau_k = k\ln((k+1)/k)$ 
(Eq.~(\ref{tau2max-av})). The inset shows the same plot with $\lambda \tau$ range
(0,5).
}
\end{figure}

For a general $k$, getting explicit solutions for $\alpha_u$ is not easy. But it is possible to
get an expression for $\tau_{2max}^{'}$. We
use the condition that for $\alpha_u=\alpha_l=1$, Eq.~(\ref{kth}) has two 
degenerate solutions $\mu=1$. This gives the condition $b_0 = 1+k- kc$. 
Simplifying, we get
\begin{eqnarray}
 \lambda \tau_{2max} & = & k \lambda \tau + \lambda \tau_{2max}^{'}
\nonumber \\
& = & k \lambda \tau - \ln(k+1) - \ln(1-e^{-\lambda \tau})
\label{tau2max-k}
\end{eqnarray}
For $k=0$, this equation reduces to Eq.~(\ref{tau2max-0}) and for $k=1$
it reduces to Eq.~(\ref{tau2max-1}). Fig.~\ref{tau-tau2max} shows the 
plot of $\lambda \tau_{2max}$ as a function of $\tau$ for different $k$
values. For each $k$ the plot is for the range $(\tau_k,\tau_{k+1})$
where $\tau_k$ is defined by $k \tau_k = \tau_{2max}$ and from Eq.~(\ref{tau2max-k}) we get $\tau_k = \ln((k+1)/k)$.

\subsubsection{$\tau_2 = n\tau$}
In this case it is not possible to obtain the stability range for 
the synchronized state in terms of $\alpha$. However, it is possible to 
obtain an explicit expression for $\tau_{2max}$ as (The detailed 
calculations are given in Appendix \ref{AppB}.)
\begin{equation}
\lambda \tau_{2max} = \lambda \tau / ( exp(\lambda \tau) - 1) 
\label{tau2max-av}
\end{equation}
The dashed line in Fig.~(\ref{tau-tau2max}) corresponds to Eq.~(\ref{tau2max-av}). It gives the correct $\tau_{2max}$ only for $\tau_2 = n\tau$. 

\section{\label{sec3}NUMERICAL ANALYSIS}

We choose two standard systems, R\"ossler and Lorenz, to confirm the main results obtained in the previous section. 

In the case of R\"ossler system given in Eq.~(\ref{rossler}), two identical systems are coupled in the drive response mode via the coupling scheme explained in Eq.~(\ref{system_def}). 
Starting from random initial conditions and choosing the system parameters in the chaotic region ($a= 0.15, b=0.2$ and $c= 10.0$),  they are evolved for 200000 units with a time step of 0.01. The correlation coefficient  $C=<y_1(t)x_1(t+\tau_2)>/\sqrt{<x_1^2(t)><y_1^2(t)>}$ between $x_1(t)$ and $y_1(t)$ shifted by  the effective $\tau_2 = | \tau_2-\tau_1|$ (hereafter referred to as $\tau_2$ itself) is calculated using the last 5000 values. The region of stability of the synchronized state  is isolated as the region where $C= 0.99$ and boundaries of stability fixed when $C$ goes below this value.

\begin{figure}
\includegraphics[width=0.9\columnwidth]{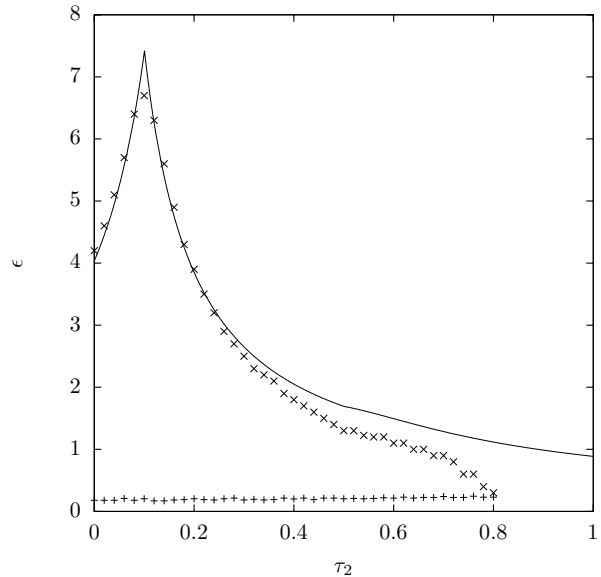}%
\caption{\label{tau2-c-ros}The limits of stability of the synchronized state of two chaotic R\"ossler systems in the parameter plane $\tau_2-\epsilon$. The solid line is the limits obtained from the stability analysis for $\lambda=0.65$. The agreement is good for values of $\epsilon > 3.0$. 
}
\end{figure}

Taking $\tau = 0.5 $, $\tau_2 $ is varied from 0 to 1.0 units in steps of 0.01. For each value of $\tau_2$, the coupling strength $\epsilon $ is increased in steps of 0.005. The appropriately shifted correlation coefficient is calculated and using the criterion mentioned above the lower and upper limits of stability are found out. The results are plotted in the parameter plane $\tau_2-\epsilon$ in Fig.~\ref{tau2-c-ros}. The overall behavior agrees with the theoretical analysis carried out in the previous section. The upper limits obtained by the stability analysis given in  Eq.~(\ref{alphau1}), Eq.~(\ref{alphau2}) and Eq.~(\ref{alphau-k1}) for the different relative ranges of $\tau_2$ are calculated for a typical value of $\lambda=0.65$ and shown as solid line. For values of $\tau_2<0.3$ or $\epsilon>3.0$, the agreement is good although for lower values, there is deviation.
For lower values of $\epsilon$ we need larger values of $\lambda$ to obtain a better fit (not shown in the figure).

\begin{figure}
\includegraphics[width=0.9\columnwidth]{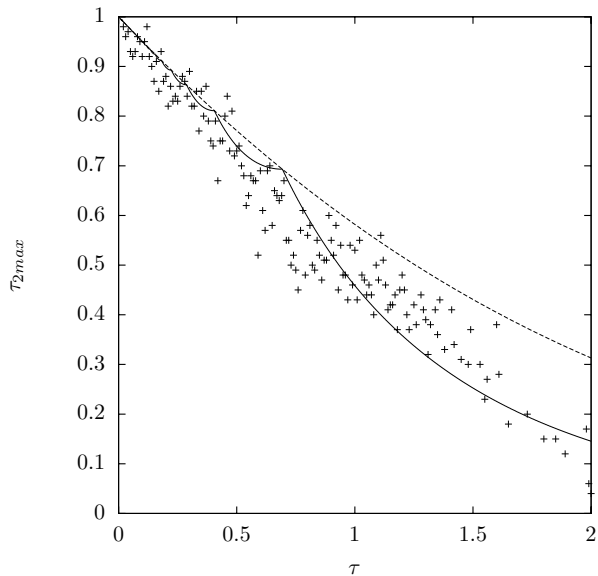}%
\caption{\label{t2m-t-ros}The maximum values of $\tau_2$ for which anticipatory synchronization is stable in two coupled R\"ossler systems is shown as a function of the rest time $\tau$. The solid and dotted lines are the similar values from theory reproduced from Fig.~\ref{tau-tau2max} for comparison. 
}
\end{figure}

By fixing the coupling $\epsilon =0.8$, we vary the reset time $\tau$ in the range (0,2.0) in steps of .01 and in each case the maximum value of $\tau_2$ for stability of synchronization is calculated using the same criterion. The results are shown in Fig.~\ref{t2m-t-ros}. The $\tau_{2max}$ obtained from theory and shown in Fig.~\ref{tau-tau2max} are reproduced here for comparison. The numerical values are found to support the results of the theoretical analysis very well. We note that here the $\lambda$ dependence cancels out and hence the agreement with the theory is much better than
that for the $\tau_2-\epsilon$ plots. 

We consider next two Lorenz systems, where the x-system is given by 

\begin{eqnarray}
	\dot{x_1} & = & a (x_2- x_1) \nonumber \\
	\dot{x_2} & = & c x_1 -x_2-x_1 x_3
\label{lorenz} \\
	\dot{x_3} & = & -b x_3 + x_1 x_2 \nonumber
\end{eqnarray} 

\begin{figure}
\includegraphics[width=0.9\columnwidth]{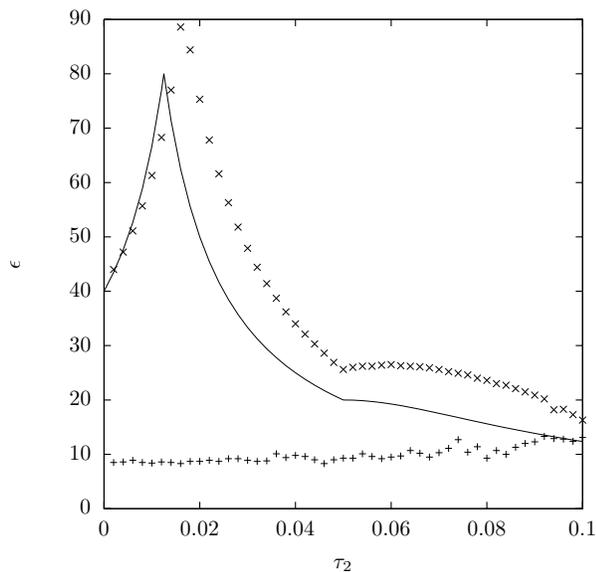}%
\caption{\label{tau2-c-lor}The maximum values of coupling $\epsilon $ for two coupled Lorenz systems as a function of the anticipatory time $\tau_2$. The solid curve is the limiting curve from theory same as in Fig.~\ref{tau2-ep} for a value of $\lambda=0.0$. 
}
\end{figure}

This is coupled to an identical y-system using the same scheme. Choosing parameter values for chaotic Lorenz as $a= 10.0, b= 8/3$ and $c= 28.0$, the analysis is repeated as in the case of R\"ossler. Here the time step chosen is 0.001 and $\tau=0.05$. The $\tau_2$ values are varied in the range (0,0.1) and the stability limits of $\epsilon$ isolated. The results are given in Fig.~\ref{tau2-c-lor}. The general behavior agrees with the theory in this case also. However, the nearest fit (shown in solid line ) is obtained for $\lambda=0.0$.

\section{\label{sec4}CONCLUSION}

We introduce a new coupling scheme with varying time delay for synchronization of two systems with delay or anticipation. The scheme has the advantage that synchronization can be  achieved with intermittent information from the driver in intervals of reset that can be pre-fixed. This also makes a detailed stability analysis analytically possible because the error dynamics becomes discrete. By assuming an average effective Lyapunov exponent $\lambda$, the stability regions and limits of stability in the parameters of coupling strength and anticipation time are worked out for specific cases. We demonstrate the method by numerical simulations in two standard systems, R\"ossler and Lorenz. The general features of the stability
region in parameter space match with the theoretical stability analysis, but more precise matching with the numerical data is not possible. This is understandable since in the analytical calculations $f^{'}$ is replaced by an effective 
$\lambda$ and
also coupling in all components of $x$ and $y$ is assumed while in numerical calculations only one component is coupled. The agreement between the theory and numerical data is reasonably good for the $\tau_{2max} - \tau$ plot,
since the $\lambda$ dependence cancels out.

The availability of three new time scales in the dynamics is suggestive of potent applications especially in secure communication. We propose that this technique will be especially successful with a bichannel transmission \cite {bocc} where one channel, that is part of the state space of the chaotic transmitter (driver), is used to synchronize  with the receiver (response) and the other forms the message along with the chaotic signal from a different part of the state space of the driver. Here since the encrypted information or cipher text is not used as the   synchronizing signal, it can be made really complex and secure. In this context our method of synchronization has the definite advantage that the synchronization channel need be transmitted only at intervals fixed by the reset time which itself forms part of the key space. This leads to bandwidth savings and requirement of noise free channel for short times at intervals. Moreover, the enhancement in the dimensionality of the key space leads to increase in security. The stability analysis reported in this paper along with the numerical simulations for standard systems helps to fix the accessible regions of the key space for better key management. This is being worked out and will be published elsewhere.

\begin{acknowledgments}

One of the authors, GA, acknowledges the hospitality and facilities at Physical Research Laboratory, Ahmedabad during the visit under associateship. 
\end{acknowledgments}

\appendix
\section{\label{AppA}Case $0 \leq \tau_2 \leq \tau$}
Here we analyze the eigenvalues $\mu_{\pm}$ of the
map~(\ref{2dmap}) given by Eq.~(\ref{ev-a}) to obtain the stability conditions. These stability conditions are shown in Fig.~\ref{tau2-ep}.
The synchronized state, $\Delta=0$, is stable if both the eigenvalues
satisfy $|\mu_{\pm}| < 1$.

\subsection{$\tau_2=0$}
\begin{figure}
\includegraphics[width=0.9\columnwidth]{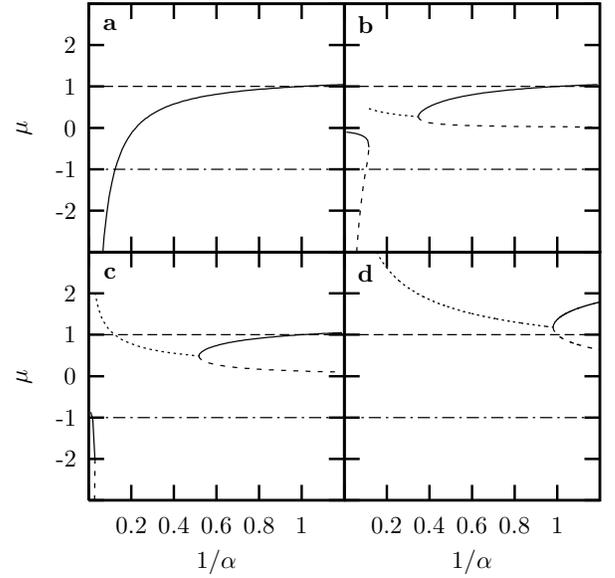}%
\caption{\label{mu-alpha}This figure shows the eigenvalues $\mu$ as a function of $1/\alpha$. (a) $\lambda \tau = 0.25$ and $\lambda \tau_2 =0$.
Here, $\mu = \alpha \left( 1 - (1-\frac{1}{\alpha}) e^{\lambda \tau} \right)$ (see Eq.~(\ref{1dmap})). (b) $\lambda \tau = 0.25$ and $\lambda \tau_2 =0.02$. The largest $\mu$ (solid line) starts from a value greater than one for $1/\alpha >1$, crosses 1 at $1/\alpha=1$, and continues till it meets the dashed line from
below (Here $a^2-4b=0$). Then $\mu$ becomes complex and the dotted line 
shows the 
magnitude $|\mu|$. This continues till we have $a^2-4b=0$ again. This point is just above the meeting
point of solid and dashed lines on the negative side. 
For smaller values of $1/\alpha$, $\mu$ again become real (but now negative) and the largest $\mu$ in magnitude jumps to the dashed line bellow. Hence the stability range is from 
$\alpha=1$ till the point where $\mu=-1$. (c) $\lambda \tau = 0.25$ and $\lambda \tau_2 =0.1$.
This figure is similar to (b), but here the dotted line ($\mu$ complex) crosses the 
magnitude
one before jumping to the negative value. Hence the stability range is now
from $\alpha=1$ till the point where the dotted line crosses one or the complex $\mu$ has magnitude one.
Crossover from the behavior (b) to (c) occurs at the peak value, $\alpha_p$, as seen
in Figure~\ref{tau2-ep}. (d) $\lambda \tau = 1.0$ and $\lambda \tau_2 =0.7$.
Here, $\tau_2 > \tau_{2max}=0.458\ldots$. Hence, the largest $|\mu|$ is always greater than one.
}
\end{figure}

For this case $b=0$. Hence, 2-d map in Eq.~(\ref{2dmap})
becomes a 1-d map given by
\begin{equation}
 \Delta_{m+1} =  \mu \Delta_m
\label{1dmap}
\end{equation}
where $\mu =\alpha \left( 1 - (1-\frac{1}{\alpha}) e^{\lambda \tau} \right)$.
Fig.~\ref{mu-alpha}a shows $\mu$ as a function of $1/\alpha$.
The fixed point $\Delta=0$ is stable provided $|\mu|<1$.
This gives the following limits on $\alpha$ for the stability of the
synchronized state.
\begin{equation}
 1 < \alpha < \frac{1+e^{-\lambda \tau}}{1-e^{-\lambda \tau}}
\label{alpha-0}
\end{equation}

\subsection{$0<\tau_2 <\tau$}

For $0<\tau_2<\tau$, the eigenvalues $\mu_{\pm}$ (Eq.~(\ref{ev-a})) display 
a rich behavior. Three different scenarios are possible. These are shown in Figs.~\ref{mu-alpha}b, \ref{mu-alpha}c and \ref{mu-alpha}d which show $\mu$ as a function of
$1/\alpha$. To determine the limits of stability of the solution $\Delta=0$ we consider the following case.

\subsubsection{$\mu = 1$ ($a^2-4b >0, a>0$)}
Putting $\mu = 1$ in Eq.~(\ref{ev-a}), we get $2 = a \pm \sqrt{a^2-4b}$. This reduces to
\begin{equation}
 1 = a-b
\end{equation}
Using the expressions (\ref{ab}) for $a$ and $b$, we get the lower limit on the stability as
\begin{equation}
 \alpha_l =1.
\end{equation}

\subsubsection{$\mu=-1$ ($a^2-4b >0, a <0$)}
Putting $\mu=-1$ in Eq.~(\ref{ev-a}), we get $-2 = a \pm  \sqrt{a^2-4b}$ which reduces to
\begin{equation}
 1+a+b=0
\end{equation}
Using the expressions (\ref{ab}), we get
\begin{equation}
 \alpha_u = \frac{e^{\lambda \tau}+1}{2e^{\lambda (\tau- \tau_2)} - e^{\lambda \tau} - 1}
\label{alpha-case2}
\end{equation}
The above expression gives the upper limit of stability for smaller values of $\tau_2$. For larger values of $\tau_2$, we use the condition $|\mu|=1$ which is considered in the next subsection.

\subsubsection{$|\mu|=1$ ($a^2-4b <0$, $\mu$ complex)}
Putting $|\mu|=1$ ($\mu$ complex) in Eq.~(\ref{ev-a}), we get $1 = \frac{1}{2} \sqrt{a^2-(a^2-4b)}$ which reduces to
\begin{equation}
 b=1
\end{equation}
Substituting from Eq.~(\ref{b}), we get the upper limit on the stability as
\begin{equation}
 \alpha_u = \frac{e^{-\lambda \tau}}{1-e^{-\lambda \tau_2}}
\label{alpha-case3}
\end{equation}
The above expression can also be used to determine the maximum $\tau_{2max}$ for a given $\tau$. This happens when there is always an eigenvalue with magnitude greater than one, i.e. when $\alpha_l = \alpha_u =1$. From Eq.~(\ref{alpha-case3}) we get the following expression.
\begin{equation}
 \lambda \tau_{2max} = -\ln (1-e^{-\lambda \tau}).
\label{tau2max-0} 
\end{equation}
Note that $\alpha_u=1$ in Eq.~(\ref{alpha-case2}) gives the same $\tau_{2max}$ 
as in Eq.~(\ref{tau2max-0}). Fig.~(\ref{mu-alpha}d) shows $\mu$
as a function of $1/\alpha$ for $\tau_2 > \tau_{2max}$ where the synchronized state is not stable.

\subsubsection{Peak}
The peak value $\alpha_p$ is given by the intersection of Eqs.~(\ref{alpha-case2}) and~(\ref{alpha-case3}) 
and leads to the conditions
\begin{equation}
 b=1, \; \; \textrm{and} \; \; a+2=0.
\end{equation}
From $b=1$, we have
\begin{equation}
 e^{-\lambda \tau_2} = 1 - \frac{1}{\alpha} e^{-\lambda \tau}
\label{peak}
\end{equation}
Substituting this in $a+2=0$, we get
\begin{equation}
 \alpha_p = \frac{e^{\lambda \tau}+3}{e^{\lambda \tau} - 1}
\label{peak-alpha}
\end{equation}
The corresponding $\tau_2$ value is given by,
\begin{equation}
 \lambda \tau_{2p} = \lambda \tau + \ln(e^{\lambda \tau}+3) - 2\ln(e^{\lambda \tau}+1)
\label{peak-tau2}
\end{equation}
Eliminating $\tau$ from Eqs.~(\ref{peak-alpha}) and~(\ref{peak-tau2}) gives Eq.~(\ref{tau2p-alphap}).

\subsection{$\tau_2 = \tau$}
This is a simple case where $a$ and $b$ in Eqs.~(\ref{ab}) reduce to
$a = c = e^{\lambda \tau}, \; \;
 b = d = \alpha (e^{\lambda \tau}- 1)$
The cases Appendix~$A2a$ and~$A2c$ in the above subsection are applicable and hence the stability 
condition for $\Delta=0$ is 
\begin{equation}
 1 < \alpha < \frac{1}{e^{\lambda \tau} - 1}
\end{equation}

\section{\label{AppB}Case $\tau_2 = n \tau$}

This corresponds to the case $\tau_2^{'}=0$ in Section IIIB. Using 
Eq.~(\ref{rec-tau2'}), we get the following recursion relation (note that n=k+1 gives the correct correspondence)
\begin{subequations}
\begin{eqnarray}
 \Delta_{m+n+1} & = & e^{\lambda \tau} \Delta_{m+n} - \alpha (e^{\lambda \tau} -1) \Delta_m \\
& = & c \Delta_{m+n} - d \Delta_m
\label{rec-ntau} 
\end{eqnarray}
\end{subequations}
where $c=e^{\lambda \tau}$ and $d=\alpha (e^{\lambda \tau} -1)$.
This leads to an $(n+1)$ dimensional map. This map can also be obtained directly from
the solution (\ref{sol-Cm}) noting that for $t=m\tau$ and $t=(m+1)\tau$ we get $\Delta_{m+n}$ and $\Delta_{m+n+1}$ respectively. In matrix form
\begin{equation}
 \left( \begin{array}{c}
 \Delta_{n+1} \\ \Delta_n \\ \Delta_{n-1} \\ \vdots \\ \Delta_1
\end{array} \right) = 
\left( \begin{array}{ccccc}
c & 0 & \cdots & 0 & - d \\ 
1 & 0 & \cdots & 0 & 0 \\ 
0 & 1 & \cdots & 0 & 0 \\
\vdots & \vdots & \vdots & \vdots & \vdots \\
0 & 0 & \cdots & 1 & 0
\end{array} \right)
\left( \begin{array}{c}
 \Delta_n \\ \Delta_{n-1} \\ \Delta_{n-2} \\ \vdots \\ \Delta_0
\end{array} \right)
\label{ndmap}
\end{equation}
The eigenvalue equation is
\begin{eqnarray}
\mu^{n+1} - c \mu^n +d & =& 0,
\label{mu-n}
\end{eqnarray}
where $n\geq 1$. 

The following general conclusions can be arrived at using
Ger\v sgorin discs. There is one disc with center at $c$ and 
radius $d$ and
$n$ discs with center at $0$ and radius $1$. All the eigenvalues 
lie within these discs. For $\alpha <1$, $c>d+1$. Hence, the disc with center at $c$ is disjoint from the other discs.
Thus one root which lies in this disc, must always have magnitude greater than one. Hence, the lower
limit of stability is $\alpha_l=1$. 

For $n=1$, Eq.~(\ref{mu-n}) becomes a quadratic equation. This is
discussed in Appendix A3.

\subsection{$n=2$}
Eq.~(\ref{mu-n}) becomes a cubic equation.
\begin{equation}
\mu^3 - c \mu^2 +d =0
\label{mu-3}
\end{equation}
At the upper stability limit, $\mu$ is complex with $|\mu|=1$. Thus, $\mu^2 - 2 \cos(\theta) \mu +1$ is a factor where $\mu = e^{\pm i \theta}$.
Using this condition we get the relation
\begin{equation}
 d^2 + cd = 1
\end{equation}
This gives a quadratic equation in $\alpha$.
\begin{equation}
 (e^{\lambda \tau} - 1)^2 \alpha^2 + e^{\lambda \tau}(e^{\lambda \tau} - 1)
\alpha -1 = 0.
\end{equation}
Using the correct solution 
the stability range is
\begin{equation}
 1 < \alpha < \frac{1}{2(e^{\lambda \tau} - 1)}\left(\sqrt{e^{2\lambda \tau}+4} - e^{\lambda \tau}\right)
\end{equation}

The maximum $\tau_{2max}$ is obtained in the next subsection.

\subsection{Any $n$}
For a general $n$ it is not possible to obtain the stability range for
the synchronized solution.

It is easy to see that the maximum $\tau_{2max}$ is obtained
if there are two degenerate eigenvalues of Eq.~(\ref{mu-n}) equal to one at $\alpha=1$.
This is possible if $c = 1/n$, $d = 1+1/n$.
Using the explicit form of $c$ or $d$ and $n=\tau_{2max}/\tau$, we get Eq.~(\ref{tau2max-av}) for
$\tau_{2max}$.

\section{\label{AppC} Negative $\lambda$}

If Lyapunov exponent $\lambda$ is negative, then Eq.~(\ref{linear_problem}) can be written as
\begin{equation}
 \dot{\Delta} = -|\lambda| \Delta - \epsilon \sum_{m=0}^{\infty}  \chi_{(m\tau,(m+1)\tau)} \Delta_m
\label{linear_problem_neg}
\end{equation}
The analysis is similar to that for positive $\lambda$. Here, we summarize
the results.

\subsection{$0 \leq \tau_{20} \leq \tau$}
For $0 \leq \tau_{20} \leq \tau$, Eq.~(\ref{linear_problem_neg}) leads to 
the recursion relation (see Eq.~(\ref{rec-tau2}))
\begin{eqnarray}
 \Delta_{m+1} 
 & = & a \Delta_m - b \Delta_{m-1} 
\label{rec-tau2-b-neg}
\end{eqnarray}
where
\begin{subequations}
\label{ab-neg}
\begin{eqnarray}
a & = & -\alpha (e^{|\lambda|(\tau_2-\tau)}-1) + e^{-|\lambda| \tau}, 
\label{a-neg} \\
b & = & \alpha e^{-|\lambda| \tau}(e^{|\lambda| \tau_2}-1)
\label{b-neg}
\end{eqnarray}
\end{subequations}
where we define
$\alpha = \epsilon/|\lambda|$ as the normalized dimensionless coupling constant.

\subsubsection{$\tau_2=0$}
For $\tau_2=0$, $b=0$. The stability limits for the synchronized state are
(see Eq.~(\ref{alpha-0}))
\begin{equation}
 -1 < \alpha < \frac{1+e^{-|\lambda| \tau}}{1-e^{-|\lambda| \tau}}
\label{alpha-0-neg}
\end{equation}

\subsubsection{$0 < \tau_2 \leq \tau$}
In this case Eq ~(\ref{rec-tau2-b-neg}) leads to a 2-d map as for the positive
$\lambda$. The eigenvalue equation and the solutions are same as Eqs.~(\ref{eq-ev-2d}) and~(\ref{ev-a}) with $a$ and $b$ defined by
Eqs.~(\ref{ab-neg}).

The lower stability limit is always $\alpha_l=-1$. For smaller values of $\tau_2$ ($\leq \tau_{2p}$), the upper limit
of stability is given by (see Eq.~(\ref{alphau1}))
\begin{equation}
 \alpha_u = \frac{1+e^{-|\lambda| \tau}}{1+e^{-|\lambda| \tau}-2e^{|\lambda| (\tau_2- \tau)}}
\label{alphau1-neg}
\end{equation}
while for larger values of $\tau_2$ ($\tau_{2p} \leq \tau_2 \leq \tau$) it is given by (Eq.~(\ref{alphau2}))
\begin{equation}
 \alpha_u = \frac{e^{|\lambda| \tau}}{e^{|\lambda| \tau_2}-1}
\label{alphau2-neg}
\end{equation}
It is interesting to note that for very large values of the
coupling constant the synchronized state is unstable. 
The maximum value of $\alpha_p$ is given by the intersection of the two curves~(\ref{alphau1-neg}) and~(\ref{alphau2-neg}).
\begin{equation}
 \alpha_p = \frac{3 e^{\lambda \tau}+1}{e^{\lambda \tau} - 1}
\label{alphap-neg}
\end{equation}
The corresponding $\tau_{2p}$ value is determined by the relation
\begin{equation}
|\lambda| \tau_{2p} = |\lambda| \tau + 2 \ln(1+e^{-|\lambda| \tau}) - \ln(3+e^{-|\lambda| \tau})
\label{tau2p-neg}
\end{equation}

\begin{figure}
\includegraphics[width=0.9\columnwidth]{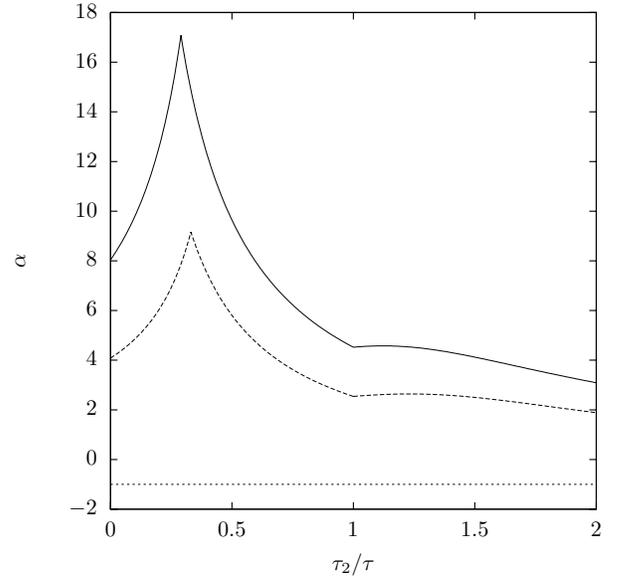}%
\caption{\label{tau2-ep-neg}The stability region
of the synchronized state
in the $|\lambda| \tau_2 - \alpha$ plane. The solid line is for  $\lambda \tau = - 0.25<0$ and the dashed line is for $\lambda \tau = -0.5$.
The lower limit of stability is 
$\alpha_l=-1$ (dotted line). For smaller values of $\tau_2 \leq \tau_{2p}$, the upper limit
of stability is given by (Eq.~(\ref{alphau1-neg}))
while for larger values of $\tau_2$ ($\tau_{2p} \leq \tau_2 \leq \tau$) it is given by (Eq.~(\ref{alphau2-neg})). The peak values are ($0.072\ldots/0.25\ldots=0.29 ,17.083\ldots$) for $\lambda\tau=-0.25$ and ($0.165\ldots/0.5=0.33\ldots, 9.16\ldots$)for $\lambda\tau=-0.5$ (see Eqs.~(\ref{alphap-neg}) and~(\ref{tau2p-neg})). For $\tau \leq \tau_2 \leq 2 \tau$ the upper limit of stability is given by Eq.~(\ref{alphau-k1-neg}).
We note that the stability limits have a similar behavior to that of Fig.~\ref{tau2-ep} for positive $\lambda$. 
}
\end{figure}

\subsubsection{$\tau_2 = \tau$}
For $\tau_2 = \tau$, the stability range is
\begin{equation}
 -1 < \alpha < \frac{1}{1-e^{-|\lambda| \tau}}
\end{equation}

\subsection{$\tau_2 > \tau$}
Let $\tau_2 = k\tau + \tau_2^{'}, \; k = 0,1,\ldots$ where $\tau_2^{'} < \tau$ as for the case of positive $\lambda$. Eq.~(\ref{linear_problem_neg}) leads to
a map of dimension $k+2$. The eigenvalue 
equation is 
\begin{equation}
\mu^{k+2} - c \mu^{k+1} + b_1 \mu + b_0 = 0
\label{kth-neg}
\end{equation}
where $c=e^{-|\lambda| \tau}$, $b_1=\alpha (1-e^{|\lambda|(\tau_2^{'}-\tau)})$
and $b_0=\alpha e^{-|\lambda| \tau}(e^{|\lambda| \tau_2^{'}}-1)$. 
For $k=0$, we recover the case $0 < \tau_2 \leq \tau$.

For $k=1, \; {\textrm i.e.} \; \tau \leq \tau_2 \leq 2\tau$,
we have a 3d-map. The lower stability limit is $\alpha_l=-1$.
The upper stability limit is
\begin{equation}
 \alpha_u = \frac{1}{2 a_2}(a_1 + \sqrt{a_1^2 +4 a_2})
\label{alphau-k1-neg}
\end{equation}
where $a_1 = e^{-2 |\lambda| \tau} ( 1 - e^{|\lambda| \tau_2^{'}}) + e^{|\lambda|(\tau_2^{'}-\tau)} -1$ and $a_2 = e^{-2 |\lambda| \tau}(e^{|\lambda| \tau_2^{'}}-1)^2$. The stability limits are plotted in Fig.~\ref{tau2-ep-neg}.

\begin{figure}
\includegraphics[width=0.9\columnwidth]{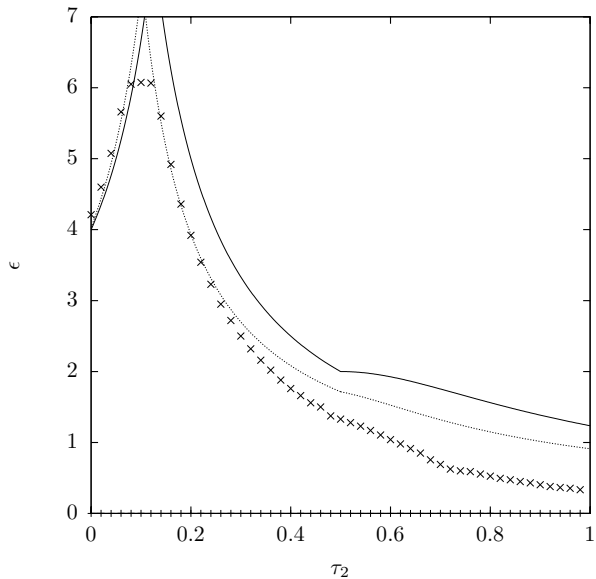}%
\caption{\label{tau2-c-ros-p}The maximum values of coupling $\epsilon $ for two coupled R\"ossler systems in the periodic region. The solid curve is for the  values from theory reproduced from Fig.~\ref{tau2-ep-neg} for a value of $\lambda=0.0$. The agreement is better with the curves in Fig.~\ref{tau2-ep} (dotted line) for a value of $\lambda=0.6$.
}
\end{figure}

We have done numerical analysis for negative $\lambda$ using two R\"ossler systems in the periodic region for $c= 2.2$. The stability limits for synchronization in the $\tau_2-\epsilon$ plane in this case is given in Fig.~\ref{tau2-c-ros-p}. The solid line is the curve from theory with $\lambda=0.0$ . The behavior of the numerical results in general agrees with the theoretical analysis. However, exact fit is not obtained for any negative $\lambda$. Surprisingly, the fit is better for positive $\lambda$ with  equations Eq.~(\ref{alphau1}) and Eq.~(\ref{alphau2}) 
(dotted line). The reason for this behavior is not clear.

\subsection{$\tau_{2max}$}
The condition for obtaining the maximum value $\tau_{2max}$ is that
$\alpha_l=\alpha_u$. For negative $\lambda$, we have $\alpha_l=-1$ and
$\alpha_u$ always remains positive. Hence, unlike the case of positive $\lambda$, the condition for obtaining
$\tau_{2max}$ is never satisfied and synchronized state is possible for any $\tau_2$ or $\tau_{2max}$ is infinite. 

\subsection{$\tau_2 = n\tau$}

This corresponds to the case $\tau_2^{'}=0$ of Appendix C2. The eigenvalue equation is (see Eq.~(\ref{mu-n}))
\begin{eqnarray}
\mu^{n+1} - c \mu^n +d & =& 0,
\label{mu-n-neg}
\end{eqnarray}
where $n\geq 1$ and $c=e^{-|\lambda| \tau}$ and $d=\alpha (1-e^{-|\lambda| \tau})$.

The following general conclusions can be arrived at using
the Ger\v sgorin discs. There is one disc with center at $c$ and 
radius $|d|$ and
$n$ discs with center at $0$ and radius $1$. All the eigenvalues 
lie within these discs. For $\alpha <1$, $d<(1-c)$. Since $c<1$, the disc with center at $c$ lies within the circle $|\mu|=1$. Hence, all the roots of Eq.~(\ref{mu-n-neg}) have magnitude less than one and the 
synchronized state is stable. Thus for any $n$ there will be range of
$\alpha$ values for which the synchronized state is stable. This supports
the conclusion reached in the previous subsection (Appendix C3) that $\tau_{2max}$ is
infinite.

For $n=1$, Eq.~(\ref{mu-n-neg}) becomes a quadratic equation. This is
discussed in Appendix C1.

For $n=2$ we have a cubic equation.
The stability range is
\begin{equation}
-1 < \alpha < \frac{1}{2(1-e^{-|\lambda| \tau})}\left(\sqrt{e^{-2|\lambda| \tau}+4} - e^{-|\lambda| \tau}\right)
\end{equation}

\end{document}